\documentstyle[12pt]{article}

\newcommand{\bz}{{\hbox{\bf Z}}}

\newcommand \qint[1]{\left[ {#1} \right]}
\newcommand \fra[2]{\displaystyle
{\frac{\textstyle {#1}}{\textstyle {#2}}}}
\newcommand \vt{\vartheta}
\begin{document}

\begin{flushright}
  Apr. 1994
\end{flushright}
\vspace{24pt}
\begin{center}
\begin{large}
{\bf A trial to find an elliptic quantum algebra for $sl_2$
using the Heisenberg and Clifford algebra}
\end{large}

\vspace{36pt}
Jun'ichi Shiraishi

\vspace{6pt}

{\it Department of physics, University of Tokyo }\\
{\it Bunkyo-ku, Tokyo 113, Japan }

\vspace{48pt}

\underline{ABSTRACT}

\end{center}

\vspace{4cm}

A Heisenberg-Clifford realization of a deformed $U(sl_{2})$ by two parameters
$p$ and $q$ is discussed. The commutation relations for this
deformed algebra have interesting connection with the theta functions.

\vfill
\newpage

{\bf  Introduction} \qquad Recently it is found in the
paper \cite{FIJKMY} that
the so called $RLL=LLR$ formalism successfully leads an elliptic
quantum algebra for $\widehat{sl}_2$ which enables us to treat the
eight-vertex model by using the infinit dimensional algebra technique
as we could do for the six-vertex model (see references in \cite{FIJKMY}).

At this stage it seems very hard to work directly with the elliptic algebra
mainly because we do not have its free field realization.
The commutation relations among the $L$ operators for the elliptic algebra
are so complicated that it's not an easy task to find
the right way to bosonize the algebra. So, we need to start from a
much simple toy model and study its bosonization.

We will discuss in the following about an two parameter deformation of the
algebra $U(sl_{2})$ using the Heisinberg algebra $[\partial_z,z]=1$
and the Clifford algebra ${\cal O}_{2n+1} {\cal O}_{2n+1} = 1,$
${\cal O}_{2m+1} {\cal O}_{2n+1} = -{\cal O}_{2n+1} {\cal O}_{2m+1}$
for $m,n \in \bz_{\geq 0}$ and
$m \neq n$.

\vspace{1cm}

{\bf  an attempt at obtaining an elliptic $sl_2$ algebra using the Heisenberg
algebra} \qquad
As a toy model for the elliptic algebra introduced in \cite{FIJKMY},
we want to discuss the following rather simple commutation relation.
Let us introduce two parameters $p, q$ and define
the $p,q$ deformation of the
universal envelopping algebra for $sl_2$ as an associative algebra
generated by $e,f,$ and $t^{\pm 1}$
having the following commutation relations

{\bf relations:}
\begin{eqnarray}
&&tet^{-1}=q^{2}e,\quad tft^{-1}=q^{-2}f, \nonumber \\
&&[e,f]= \displaystyle{\sum_{n \in \bz}} (-1)^{n} t^{2n+1} p^{(n+1/2)^2}.
\quad \cdot\cdot\cdot(*) \nonumber
\end{eqnarray}
We note that the form of the right hand side of the equation $(*)$ is
nothing but the theta function.

It is well known that the Lie algebra $sl_2$ and the quantum algebra
$U_q(sl_2)$ have the following Heisenberg realization \cite{AOS}

{\bf $sl_2$ case:}
\begin{eqnarray}
e=-\partial_z,
&h=-2 z \partial_z + \lambda,
&f=z^{2} \partial_z - \lambda z,  \nonumber
\end{eqnarray}

{\bf $U_q(sl_2)$ case:}
\begin{eqnarray}
e=-\frac{1}{z}\qint{\vt},
&t=q^{-2 \vt + \lambda},
&f=z \qint{ \vt - \lambda},\nonumber
\end{eqnarray}
here, we set $\vt=z\partial_z$ and $\qint{A}=(q^{A}-q^{-A})/(q-q^{-1})$.

Now we state our main result.

{\bf Proposition:}
{\it We have the following
Heisenberg-Clifford realization for our elliptic $sl_2$ algebra, }

\begin{eqnarray}
&&e=-\frac{1}{z}\sum_{n \in \bz} a_{2n+1}
q^{(2n+1)\vt} p^{(n+1/2)^2/2}{\cal O}_{|2n+1|},   \nonumber\\
&&t= q^{-2\vt+\lambda} ,   \nonumber\\
&&f=z\sum_{n \in \bz} b_{2n+1}q^{(2n+1)(\vt-\lambda)}p^{(n+1/2)^2/2}
{\cal O}_{|2n+1|},
   \nonumber
\end{eqnarray}
{\it where}
\begin{eqnarray}
&&a_{2n+1}=\fra{q^{-(2n+1)(\lambda+1)/2}(-1)^{-n/2}}
           {\sqrt{q^{2n+1}-q^{-2n-1}}}, \nonumber\\
&&b_{2n+1}=\fra{q^{(2n+1)(\lambda+1)/2}(-1)^{-n/2}}
           {\sqrt{q^{2n+1}-q^{-2n-1}}} \nonumber
\end{eqnarray}
{\it for $n \in \bz$.}

Proof:\quad
Let $n,m \in \bz$ and $n \neq m$.
To satisfy the commutation relation $(*)$ we must have
$$
a_{2n+1}b_{2n+1}=\fra{(-1)^{-n}}{q^{2n+1}-q^{-2n-1}},
$$
\begin{eqnarray}
&&a_{2n+1}b_{2m+1}q^{(n-m)(\lambda+1)}
 -a_{2m+1}b_{2n+1}q^{-(n-m)(\lambda+1)}=0.\nonumber
\end{eqnarray}
These equations can be solved and have the solution stated as above.
{\it QED}

\vspace{1cm}

We discussed an Heisenberg realization of a two parameter deformation
of the universal envelopping algebra of $sl_2$.
So far, we have not found a Hopf algebra structure which is nesesarry to
construct tensor product representations of our algebra.
The connection between out algebra and the algebra defined in \cite{FIJKMY}
is not clear.
We hope that to do a loop-algebraic extension of our algebra
is also possible in the similar manner as discussed in \cite{AOS}.
It is an interesting problem to think why the operators $e$ and $f$
resemble theta functions except for the q-dependent factor
when we bosonize them.
How can we interpret the infinit dimensional Clifford
algebra that is so trivial that we
did not need when $p=0$?


\vspace{1cm}

{\bf Acknowledgements} \qquad
The author would like to thak H. Awata, T. Eguchi, O. Foda, M. Jimbo,
T. Miwa and K. Sugiyama for helpful discussions.
He is grateful to kind hospitality at
YITP and RIMS.

\end{document}